\documentclass[lettersize,onecolumn]{IEEEtran}
\usepackage{amsmath,amsfonts}
\usepackage{algorithmic}
\usepackage{algorithm}
\usepackage{array}
\usepackage[caption=false,font=normalsize,labelfont=sf,textfont=sf]{subfig}
\usepackage{textcomp}
\usepackage{stfloats}
\usepackage{url}
\usepackage{verbatim}
\usepackage{graphicx}
\usepackage{cite}
\usepackage{mathtools,etoolbox}
\mathtoolsset{showonlyrefs,showmanualtags}

\newtheorem{theorem}{Theorem}
\newtheorem{lemma}{Lemma}
\newtheorem{assumption}{Assumption}
\newtheorem{corollary}{Corollary}

\newcommand\norm[1]{\left\lVert#1\right\rVert}

\newcommand{\E}{{\mathbb E}}

\begin{document}

\title{Risk Bounds on MDL Estimators for Linear Regression Models with Application to Simple ReLU 
Neural Networks}

\author{
  Yoshinari Takeishi and Jun'ichi Takeuchi\\
  Faculty of Information Science and Electrical Engineering\\
                   Kyushu University,  Fukuoka 819-0395, Japan\\
                    Email: \{takeishi,tak\}@inf.kyushu-u.ac.jp
\thanks{This work was supported by JSPS KAKENHI, Grant Number JP23H05492. This paper was 
presented in part at the 2024 IEEE International Symposium on Information
Theory.}
}

\markboth{IEEE TRANSACTIONS ON INFORMATION THEORY,~Vol.~X, No.~X, XXXX~20XX}%
{Shell \MakeLowercase{\textit{et al.}}: A Sample Article Using IEEEtran.cls for IEEE Journals}


\maketitle

\begin{abstract}
To investigate the theoretical foundations of deep learning from the viewpoint of 
the minimum description length (MDL) principle,
we analyse risk bounds of MDL estimators based on two-stage codes
for simple two-layers neural networks (NNs) with ReLU activation.
For that purpose, we propose a method to design two-stage codes 
for linear regression models
and establish an upper bound on the risk of the corresponding MDL estimators
based on the theory of MDL estimators originated by Barron and Cover (1991). 
Then, we apply this result to the simple two-layers NNs with ReLU activation which
consist of $d$ nodes in the input layer, $m$ nodes in the hidden layer and one output node. 
Since the object of estimation is only the $m$ weights from the hidden layer to the output node in our setting,
this is an example of linear regression models.
As a result, we show that the redundancy of the obtained two-stage codes
is small owing to the fact that 
the eigenvalue distribution of the Fisher information matrix of the NNs is strongly biased,
which was shown by Takeishi et al. (2023)
and has been refined in this paper.
That is, we establish a tight upper bound
on the risk of our MDL estimators. 
Note that our risk bound for the simple ReLU networks, 
of which the leading term is $O(d^2 \log n /n)$,
is independent of the number of parameters $m$.
%
%
\end{abstract}

\begin{IEEEkeywords}
Minimum Description Length Principle, Two-Stage Code, Neural Network, Fisher Information Matrix.
\end{IEEEkeywords}

\section{Introduction}
Deep Learning has achieved remarkable success in various domains, but the lack of theoretical guarantees raises concerns about the reliability and generalizability of these models. In order to address these concerns, there is a need to establish theoretical foundations that provide assurances for the learning algorithms in the deep neural networks.
We tackle this problem based on the minimum description length (MDL) principle, which suggests that compressing the obtained data with the shortest description length leads to effective learning.
The mathematical justification of the MDL estimator based on the two-stage code was clarified by Barron and Cover in 1991 \cite{BC91}.

In this paper, we focus on a simplified scenario where we consider the parameter estimation of two-layer neural networks (NN) with ReLU activation. 
The target NNs consist of $d$ nodes in the input layer, $m$ nodes in the hidden layer, and one output node. 
We assume that $m\gg d$, which is reasonable in the context of recent successful large-scale neural networks.
The object of estimation is only the $m$ weights from the hidden layer to the output node. 
When we obtain $n$ input-output pairs of the network, the problem of estimating the network's parameters from these pairs is considered within the framework of supervised learning.

The evaluation of the statistical risk of the MDL estimator by Barron and Cover \cite{BC91} 
is within the framework of unsupervised learning, but 
if we employ a code for the model description which does not depend on the inputs,
it can be applied to supervised learning.
In this paper, we consider the setting.
Simply applying the technique of \cite{BC91},
we can obtain an upper bound of the risk of the MDL estimator in the form of $O(m\log n/n)$,
which is insufficient for common cases for deep neural networks
where $m$ is large compared to $n$.

To design our two-stage code, we take an advantage of the
strong bias of the eigenvalue distribution of the Fisher information matrix (FIM) of
the target model, which was rigrously shown by Takeishi et al. \cite{Takeishi,TakeishiNN} and has been refined in this paper. 
Specifically, under certain conditions, 
the sum of the first $d(d+3)/2$ eigenvalues constitutes approximately 97\% or more of the total sum of $m$ eigenvalues.
This suggests that the important directions in learning depend only on the dimension $d$ of the input, and not on $m$. Note that $m$ must be larger than $d(d+3)/2$ at a minimum to observe this phenomenon.
Building upon this observation, we propose to construct a two-stage code by only considering the coding along these dominant eigenvalue directions.

Then, by Barron and Cover's theory \cite{BC91}, we can obtain a tight upper bound on the statistical risk of our MDL estimator in this setting (Theorem \ref{thm_redbound2}). Note that 
 the leading term of our risk bound is $O(d^2 \log n /n)$, which is independent of $m$. 

In this paper, we first review Barron and Cover's theory (Section \ref{MDLsec}). To facilitate the analysis, we begin by deriving a risk upper bound based on the exact eigenvalue decomposition for a general linear regression problem (Section \ref{linearmodel}). Following this, we derive a risk upper bound based on the approximate eigenvalue decomposition for the parameter estimation problem in simple ReLU networks (Section \ref{NNsec}).

\subsection{Related Works}

Concerning the FIM of two-layer neural networks, Chizat and Bach \cite{Chizat}
 examined
 the eigenvalue distribution in the context of the Neural Tangent Kernel (NTK) \cite{Jacot} 
 and revealed the occurrence of grouping among eigenvalues of the same magnitude. This suggests the approximate eigendecomposition 
 of the FIM shown by \cite{Takeishi,TakeishiNN}, which characterizes the eigenvalues and eigenspaces of the first three groups in detail when the activation function is ReLU.
Additionally, Iida et.al \cite{IidaISIT2022} demonstrated that even when the activation function of a two-layer neural network is 
a softplus function, similar approximate eigendecomposition holds.
Related researches on the learning of the two-layer neural networks are
summarized in Section 6 of \cite{Bartlett}.

About the MDL estimators based on two-stage codes, the theory 
originated by Barron and Cover \cite{BC91} is important,
which gives the inequality that
the risk bound of the MDL estimator is upper bounded by the redundancy of the corresponding two-stage code 
and the index of resolvability. See \cite{Grunwald} for the details and related works.
More recently, a technique to design MDL estimators without quantizing parameter spaces was developed \cite{CB2014a,CB2014b,KT2020},
but we do not discuss it in this paper.


\section{MDL Estimator and its Statistical Risk}
\label{MDLsec}
In this section, we define a two-stage code and an MDL estimator in supervised learning, and review its statistical risk derived by Barron and Cover.
Consider a parametric model $\mathcal{M}=\{p_v(y|x) : v \in V \subset \mathbb{R}^m\}$ of conditional probability densities. 
For $\mathcal{M}$ and a data string $(x^n,y^n)=\{(x^{(1)},y^{(1)}),\ldots, (x^{(n)},y^{(n)})\}$, we define 
\begin{align}
    L(y^n|x^n)=\min_{\bar{v} \in \ddot{V}_n} \Bigl( -\log p_{\bar{v}}(y^n|x^n) + L(\bar{v}) \Bigr).
\end{align}
This $ L(y^n|x^n)$ is the description length for $y^n$ given the covariate $x^n$.
Here, $\ddot{V}_n \subset V$ is a quantized version of the parameter space $V$
and $L(\bar{v})$ a code length function which satisfies Kraft's inequality
over $\ddot{V}_n$:
\begin{align}
    \sum_{\bar{v}\in\ddot{V}_n}\exp(-L(\bar{v})) \le 1,
\end{align}
where we employ `nat' for the unit, hence 
the base of the Kraft's inequality is $e$.
We call the pair $(\ddot{V}_n,L)$
as a two-stage code for the model $\mathcal{M}$.

We define an MDL estimator $\ddot{v}=\ddot{v}(x^n,y^n)$
for the parameter $v$
based on the two-stage code $(\ddot{V}_n,L)$
as 
\begin{align}
    \ddot{v}
    =
    \arg \min_{\bar{v}\in \ddot{V}_n}
    \bigl( -\log p_{\bar{v}}(y^n|x^n) + \alpha L(\bar{v})\bigr),
\end{align}
where $\alpha > 1$ is a hyper parameter of the MDL estimator.
In this form of MDL estimator, $\alpha L(\bar{v})$ with $\alpha > 1$
is used as the model description length
instead of $L(\bar{v})$.
The reason why $\alpha$ is inserted is that without $\alpha$ (if $\alpha$ were $1$),
we could not show nice upper bounds on the risk by the recipe of Barron and Cover\cite{BC91,Grunwald}.
Let 
\begin{align}
    L_\alpha(y^n|x^n) &= \min_{\bar{v}\in \ddot{V}_n}
    \bigl( -\log p_{\bar{v}}(y^n|x^n) + \alpha L(\bar{v})\bigr)\\
    &= 
     -\log p_{\ddot{v}}(y^n|x^n) + \alpha L(\ddot{v})\\
     &= -\log \bigl( p_{\ddot{v}}(y^n|x^n)e^{-\alpha L(\ddot{v})}\bigr).
\end{align}
This $L_\alpha(y^n|x^n)$ is a code length function which actually induces our MDL estimator.
Note that $L_\alpha(y^n|x^n)$ satisfies Kraft's inequality 
for each $x^n$.
Let $q(y^n|x^n)=p_{\ddot{v}}(y^n|x^n)e^{-\alpha L(\ddot{v})}$, which
 is a weak probability density 
for each $x^n$.

\subsection{Barron and Cover's Theorem}
To describe the following theorem, we introduce the conditional R\'enyi divergence 
\[
d_{\lambda}(p_1, p_2) = -\frac{1}{1 - \lambda} \log 
\mathbb{E}\left( \frac{p_2(y|x)}{p_1(y|x)} \right)^{1-\lambda},
\]
where the expectation is taken for $(x,y)\sim {p_1(y|x)p(x)}$.
The divergence $d_{\lambda}(p^*, p)$ can be regarded as a generalization error for the estimation $p$, where $p^*=p_{v^*}(y|x)$ is the true distribution. 
Let $\ddot{p}=p_{\ddot{v}}(y|x)$ denote the distribution with our MDL estimator.
The following theorem gives us an upper bound of the expectation of $d_{\lambda}(p^*, \ddot{p})$.
\begin{theorem}[Barron and Cover 1991, Li 1999, Gr\"unwald 2007]\label{thm_riskbound}
For the MDL estimator $\ddot{v}=\ddot{v}(x^n,y^n)$ based on a two-stage code with $\alpha>1$, we have   
\begin{align}
\mathbb{E} \left[d_{\lambda}(p^*, \ddot{p})\right]
\leq \frac{1}{n} \mathbb{E} \left[\log \frac{p^*(y^n|x^n)}{q(y^n|x^n)}\right]
,\label{eqn_riskbound}
\end{align}
where $\lambda\in(0,1-\alpha^{-1}]$ and the expectation is taken for $(x^n,y^n)\sim{p^*(y^n|x^n)p(x^n)}$
and $q(y^n|x^n)={p}_{\ddot{v}}(y^n|x^n)e^{-\alpha L(\ddot{v})}$.
\end{theorem}

{\it Remark 1}: Since $\ddot{v}$  depends on $x^n$ and $y^n$, $d_{\lambda}(p^*, \ddot{p})$ also depends on them. The expectation on the left side in \eqref{eqn_riskbound} is taken with respect to those $x^n$ and $y^n$.

{\it Remark 2}: This theorem is for supervised learning, which is out of scope of 
the theorems in \cite{BC91,Li1999,Grunwald} in general,
but when $L(u)$ does not depend on the covariate $x^n$, then the theorems can be applied
for supervised learning as commented in \cite{BC91}.
Noting this matter, the proof of Theorem~\ref{thm_riskbound} is seen in
 \cite{Grunwald,Li1999}. 
 A similar bound for 
 supervised learning in 
 more general setting is in \cite{KT2020}.

Proof of the theorem is stated in Appendix \ref{append-1}.

When $p^*$ belongs a model $\mathcal{M}=\{p_v(y|x) : v \in V \subset \mathbb{R}^m\}$, we can typically design a two-stage code with the quantization of $V$ is designed based on the Fisher information of $v$. 
In the following sections, we describe the specific problem settings and the design of the two-stage code for that scenario.
Then, we derive specific upper bounds for the right-hand side of \eqref{eqn_riskbound}.

\section{linear regression model}
\label{linearmodel}
Let us consider the following model in a linear regression problem. Let ${N}(\mu,\sigma^2)$ denote the Gaussian distribution with mean $\mu$ and variance $\sigma^2$.
The scalar $y\in\mathbb{R}$ is represented as
\begin{align}\label{rvmodel}
y=X v^T + \epsilon, 
\end{align}
where the row vector $X=X(x)\in \mathbb{R}^{1\times m}$ and $v\in \mathbb{R}^{1\times m}$, and the noise $\epsilon\in\mathbb{R}$ is subject to ${N}(0,\sigma^2)$. For consistency with later sections, we assumed that $X$ is a deterministic function of $x$, and that $x$ is stochastic.
In this case, the conditional density $p_{v}(y|x)$ can be expressed as 
\[
p_{v}(y|x) = N(y|X v^{T},\sigma^2).
\]

Now, let us consider the problem of estimating $v^*$ from $n$ training data $\{(x^{(1)},y^{(1)}),\ldots, (x^{(n)},y^{(n)})\}$, where we have for $t$ = 1,2,\dots,$n$
\begin{align*}
y^{(t)}&=X^{(t)} v^{*T} + \epsilon^{(t)}\\
X^{(t)}&= \varphi\left(x^{(t)} W\right),
\end{align*}
and the variables $\epsilon^{(1)},\ldots, \epsilon^{(n)}$ are independently subject to ${N}(0,\sigma^2)$.
Let $p_{v^*}(y|x)$ denote the conditional probability density of $y$ given
$x$, that is,
\[
p_{v^*}(y|x) = N(y|X v^{*T},\sigma^2).
\]
Here, we assume the parameter space $V = \{ v : \norm{v} \le 1 \}$,
that is, we assume $v^* \in V$.

In the context of parameter estimation problems, the Fisher information matrix (FIM) of the model plays a crucial role in understanding the structure of the parameter space. 
The FIM $I\in \mathbb{R}^{m\times m}$ is defined as
\begin{align*}
I_{ij} &=\mathbb{E}\left[\frac{\partial}{\partial v_i}\log p_v(y|x)
  \frac{\partial}{\partial v_j}
\log p_v(y|x) \right]\\
&= -\mathbb{E}\left[\frac{\partial^2}{\partial v_i\partial v_j}\log p_v(y|x) \right],
\end{align*}
where the expectation is taken for $(x,y)\sim 
p_v(y|x)p(x)$.
Since 
\begin{align*}
\frac{\partial^2}{\partial v_i \partial v_j}
\log p_v(y|x) =
\frac{\partial^2}{\partial v_i \partial v_j}
\Bigl( -\frac{(y-Xv^T)^2}{2\sigma^2}\Bigr)
= -\frac{X_i X_j}{\sigma^2},
\end{align*}
we have 
\begin{align*}
I_{ij} =\frac{\mathbb{E}\left[X_i X_j \right]}{\sigma^2}=\frac{J_{ij}}{\sigma^2},
\end{align*}
where we have defined the matrix $J=\mathbb{E}[X^TX]$.
Additionally, define the empirical FIM $\hat{I}\in \mathbb{R}^{m\times m}$ as 
$\hat{I}=\hat{J}/\sigma^2$, where 
\begin{align*}
\hat{J}=\frac{1}{n}\sum_{t=1}^n X^{(t) T}X^{(t)}.
\end{align*}

When the eigenvalues of $ J $ are arranged in descending order, let the $ i $-th eigenvalue be $ \mu_i $ ($ i = 1, 2, \ldots, m $). Since $ J $ is a positive semi-definite matrix, $ \mu_m \geq 0$.
Furthermore, for \( D = 1, 2, \ldots, m \), we define
\[
\beta_D = 1 - \frac{\sum_{i=1}^D \mu_i}{\mathrm{tr}(J)}.
\]
In other words, $\beta_D$ represents the proportion of the sum of the eigenvalues from the $(D+1)$-th onward to the total sum of the eigenvalues of 
$J$. Therefore, \(\beta_D\) is monotonically decreasing with respect to \( D \), and \(\beta_m = 0\).

\subsection{Risk Bound for Linear Regression}
In the following theorem, we show an upper bound on the right-hand side of \eqref{eqn_riskbound}, which is the conditional redundancy of a two-stage code.
\begin{theorem}\label{thm_redbound}
For the linear regression model $\mathcal{M}=\{p_v(y|x) : \norm{v} \leq 1\}$, we can design a two-stage code, which satisfies the following.
For each $n$ and each $D = 1, 2, \ldots, m$,
\begin{align}
\mathbb{E} \left[\log \frac{p^*(y^n|x^n)}{q(y^n|x^n)}\right]
\le
\frac{\alpha D(\log n+1)}{2}+\alpha\sum_{i\leq D}
\left(\log c_i +\frac{1}{c_i\sqrt{n}}\right)
+\frac{2n\beta_D \mathrm{tr}(J)}{\sigma^2},\label{eq_redbound}
\end{align}
where $c_i = \sqrt{\mu_i/(\alpha\sigma^2)}$.
\end{theorem}

The following corollary is immediately derived from Theorems \ref{thm_riskbound} and \ref{thm_redbound}.
\begin{corollary}\label{cor_riskbound}
Let $\alpha>1$ be a hyper parameter of the MDL estimator.
For a linear regression model $\mathcal{M}=\{p_v(y|x) : \norm{v} \leq 1\}$, we can design a two-stage code, which satisfies the following.
For each $n$ and each $D = 1, 2, \ldots, m$,
\begin{align}
\mathbb{E} \left[d_{\lambda}(p^*, \ddot{p})\right]
\leq \frac{\alpha D(\log n+1)}{2n}+\frac{\alpha}{n}\sum_{i\leq D}
\left(\log c_i +\frac{1}{c_i\sqrt{n}}\right)
+\frac{2\beta_D \mathrm{tr}(J)}{\sigma^2},
\label{eqn_riskbound_cor}
\end{align}
where $c_i = \sqrt{\mu_i/(\alpha\sigma^2)}$.
\end{corollary}

{\it Remark}: According to the assumption $\norm{v} \leq 1$ in out model $\mathcal{M}$, our estimator $\ddot{p}$ 
behaves similarly to L2 regularization. Together with the assumption $\norm{v^*} \leq 1$, we can obtain the tight upper bound on the risk for $\ddot{p}$.

If $\mathrm{tr}(J)$ is independent of $m$ and eigenvalue distribution of $J$ is strongly biased,
then the term 
$2\beta_D \mathrm{tr}(J)/\sigma^2$ is negligible, by controlling $D$.
This is because if the eigenvalue distribution is strongly biased, we can expect $\beta_D$ to be sufficiently small for some $D\ll p$.
In fact, this holds for simple ReLU networks, as discussed in the next section.

\subsection{Design of Two-Stage Code}
We will design our two-stage code for the linear regression model.
Let $u^{(i)}\in \mathbb{R}^{1\times m}$ be the unit eigenvector of the $i$-th eigenvalue $\mu_i$. Then, we have the spectral decomposition of $J$ as 
\begin{align}
    J = \sum_{i=1}^m\mu_i u^{(i)T}u^{(i)}.\label{edecomp_line}
\end{align}
We will encode only the directions of $u^{(1)},\ldots, u^{(D)}$ and ignore the other directions ($m-D$ dimensional).

Define a $m\times m$ matrix $K$ as
\begin{align}
    K = (u^{(1)T},u^{(2)T},\ldots,u^{(m)T})^T.
\end{align}
Since $K$ is an orthogonal matrix,
\begin{align}
    K^{-1}
=K^T=(u^{(1)T},u^{(2)T},\ldots,u^{(m)T})
\end{align}
holds.
Then, 
we introduce a new parameter $\theta$ by $\theta= vK^{-1}$, 
that is
\begin{align}
v = \theta K = \sum_{i=1}^m \theta_i u^{(i)},
\end{align}
which means that $\theta$ is the coordinate system with the basis 
$\{ \tilde{u}^{(i)} \}_{1\le i \le m}$.
Let 
$\bar{p}_\theta(y|x)=p_{\theta K}(y|x)$.  

For $\Theta:= \{ \theta =vK^{-1} : \norm{v} \le 1 \}=\{ \theta : \norm{\theta} \le 1 \}$, 
we design the quantized parameter space $\ddot{\Theta}_n$ as follows.
We quantize each component of $\theta$ independently.
For $\theta_i$ with $i \ge D+1$, we put just one quantized point at the origin,
for which any codeword is not necessary.
For $\theta_i$ with $i \le D$, we consider the well-known
setting of MDL frameworks for the precision of quantization:
$\Delta_i = 2\sqrt{\alpha/(\mu'_i n)}$,
where $\mu'_i$ denotes the $i$-th eigenvalue of the FIM $I$.
Recalling that $I=J/\sigma^2$, 
we employ 
\begin{align}
\Delta_i=2\sqrt{\alpha\sigma^2/(n\mu_i)}\label{delta_opt}
\end{align}
for our quantization.
Specifically, the number of quantum points for $u_i$ is defined as
$q_i=\left\lceil2/\Delta_i\right\rceil$.
Letting $q'_i=q_i-1$, the quantization space for $\theta_i$, denoted as $\ddot{\Theta}_n^{i}$, is defined as 
\begin{align*}
    \ddot{\Theta}_n^{i} = 
    \begin{cases}
        \left\{-\Delta_i q'_i/2+ \Delta_i k \ |\   k=0,1,\ldots,q'_i\right\}&  \text{if } i \leq D, \\       
        \{ 0 \} & \text{if } i > D.
    \end{cases}
\end{align*}

Next, we will determine the parameter description length.
For $i \le D$, a sufficient code length is given by
\begin{align*}
  \log q_i
    &= \log  \left\lceil\sqrt{n\mu_i/(\alpha\sigma^2)}\right\rceil.
\end{align*}
Otherwise, $0$ bit is sufficient. Hence, we can define the parameter description length 
$L_{i}$ for 
$\theta_i \in \ddot{\Theta_i}$ as
\begin{align}\label{model_dl_1}
     L_{i}(\theta_i) = 
\begin{cases}
\log  \left\lceil\sqrt{n\mu_i/(\alpha\sigma^2)}\right\rceil
& \text{if $i\leq D$}
 \\        
    0  & \text{if $i>D$}
    \end{cases}
\end{align}
Then, define the description length $L$ over $\ddot{\Theta}_n = \prod_i\ddot{\Theta}_n^i$
as 
\begin{align}\label{model_dl_2}
L(\theta)=\sum_i L_i(\theta_i).  
\end{align}
If we set $c_i = \sqrt{\mu_i/(\alpha\sigma^2)}$, then we have 
\begin{align}
  \log  \left\lceil\sqrt{n\mu_i/(\alpha\sigma^2)}\right\rceil
    &\leq \log ( c_i\sqrt{n}+1)\\
    &= (1/2)\log n+\log c_i+\log ( 1+1/(c_i\sqrt{n}))\\
    &\leq (1/2)\log n+\log c_i+1/(c_i\sqrt{n}).\label{param_length}
\end{align}

\subsection{Proof of Theorem \ref{thm_redbound}}
We evaluate the left-hand side of \eqref{eq_redbound} under the variable transformation $\theta=vK^{-1}$.
Since $-\log \bar{p}_\theta(y^n|x^n)$ is quadratic in $\theta$,
and
since its Hessian is $n K\hat{J}K^T/\sigma^2$,
we have for an arbitrary $\theta$, 
\begin{align}
    \log\frac{\bar{p}_{\theta^*}(y^n|x^n)}{\bar{p}_\theta(y^n|x^n)}
    =&
    - \left.\nabla \log \bar{p}_\theta(y^n|x^n)\right|_{\theta=\theta^*}(\theta-\theta^*)^T
    +
    \frac{n(\theta-\theta^*)K\hat{J}K^T(\theta-\theta^*)^T}{2\sigma^2},
\end{align}
where $\nabla$ denotes the gradient with respect to $\theta$, and $\theta^*=v^*K^{-1}$.
Since the expectation of 
a score function at the true parameter is zero, 
we have
\[
\mathbb{E}[\left. \nabla \log \bar{p}_\theta(y^n|x^n)\right|_{\theta=\theta^*} ]=0,
\]
which yields
\begin{align}
  \mathbb{E}\Bigl[
  \log\frac{\bar{p}_{\theta^*}(y^n|x^n)}{\bar{p}_\theta(y^n|x^n)}
\Bigr]
    =
    \frac{n(\theta-\theta^*)KJK^T(\theta-\theta^*)^T}{2\sigma^2},
\end{align}
where we have used $\mathbb{E}[\hat{J}]=J$.
Recalling the definition of $\ddot{\theta}$, we have for an arbitrary $\theta \in \ddot{\Theta}_n$
\begin{align}
  \mathbb{E}\Bigl[
  \log\frac{\bar{p}_{\theta^*}(y^n|x^n)}{\bar{p}_{\ddot{\theta}}(y^n|x^n)e^{-\alpha L(\ddot{\theta}))}}
\Bigr]&\le
  \mathbb{E}\Bigl[
  \log\frac{\bar{p}_{\theta^*}(y^n|x^n)}{\bar{p}_{\theta}(y^n|x^n)e^{-\alpha L(\theta))}}
\Bigr]\\
&=
    \frac{n(\theta-\theta^*)KJK^T(\theta-\theta^*)^T}{2\sigma^2}+\alpha L(\theta).\label{expRn}
\end{align}

By \eqref{edecomp_line} and the definition of $K$,
\begin{align}
(\theta-\theta^*)KJK^T(\theta-\theta^*)^T 
&=\sum_{i=1}^D \mu_i (\theta_i-\theta_i^*)^2+\sum_{i=D+1}^m \mu_i (\theta_i-\theta_i^*)^2\nonumber\\
&\leq \sum_{i=1}^D \mu_i (\theta_i-\theta_i^*)^2+\sum_{i=D+1}^m \mu_i\cdot 2^2\nonumber\\
&=\sum_{i=1}^D \mu_i (\theta_i-\theta_i^*)^2
+4\beta_D \mathrm{tr}(J)
\label{quadbound}
\end{align}
holds.
Due to the design of our quantization,
there exists a $\bar{\theta}_i \in \ddot{\Theta}_i$ such that
$|\bar{\theta}_i-\theta_i^*|\le \Delta_i/2$.
Then from \eqref{delta_opt},
\[
\mu_i (\bar{\theta}_i-\theta_i^*)^2 
\le \frac{\alpha\sigma^2}{n}
\]
holds.
Hence by \eqref{quadbound}, we have
\begin{align}
&(\bar{\theta}-\theta^*)KJK^T(\bar{\theta}-\theta^*)^T \leq \frac{D\alpha\sigma^2}{n}+4\beta_D \mathrm{tr}(J)
\end{align}
Together with \eqref{expRn}, we have
\begin{align}
\mathbb{E} \left[\log \frac{p^*(y^n|x^n)}{q(y^n|x^n)}\right]\le   
\frac{D\alpha}{2}+\frac{2n\beta_D \mathrm{tr}(J)}{\sigma^2}
+\alpha L(\bar{\theta}).
\end{align}
Here from \eqref{model_dl_1} and \eqref{param_length}, we derive the theorem.

\section{Simple ReLU Networks}
\label{NNsec}
Let us consider the problem of estimating the parameters of the last layer in a neural network with one hidden layer.
This problem can be considered a special case of the linear regression problem discussed in Section \ref{linearmodel},
\begin{align}
y=X v^T + \epsilon, 
\end{align}
where 
the row vector $v\in\mathbb{R}^{1\times m}$ is a 
weight vector of the last layer and the noise $\epsilon\in\mathbb{R}$ is subject to ${N}(0,\sigma^2)$.

In this setting, we assume that  
each entry of the input $x\in \mathbb{R}^d$ independently follows $N(0,1)$, 
The number of neurons in the hidden layer is $m$ and assume that $m\gg d$. 
Let $W \in \mathbb{R}^{d\times m} $ be a weight matrix of the first layer
.
Then, the output of the hidden layer $X\in \mathbb{R}^{1\times m}$ 
is represented as
\begin{align*}
X= \varphi\left(x W\right), 
\end{align*}
where $\varphi$ is the ReLU activation function and applied element-wise.
\subsection{Analysis of Fisher Information Matrix}
\label{SecFIM}
To conduct a similar discussion as in Section \ref{linearmodel}, we need the eigenvalue decomposition of the Fisher information matrix of the model. However, in large-scale neural networks, 
$m$ is extremely large, and computing the eigenvalues and eigenvectors of the matrix presents significant computational challenges.

Instead of an exact eigenvalue decomposition, the following theorem (Theorem 2 in \cite{TakeishiNN}) describes an approximate eigenvalue decomposition. 
\begin{theorem}[Takeishi et.al 2023]
\label{thm_decomp}
Assume that
all the columns of matrix $W$ are non-zero vectors.
For the matrix $J=E[X^T X]$ and $m$-dimensional row vectors $v^{(0)} $, $ \{ W_l\}_{1\leq l\leq d} $, $ \{ v^{(\gamma)}\}_{1\leq \gamma\leq d} $, $ \{ v^{(\alpha,\beta)}\}_{1\leq \alpha< \beta\leq d} $ determined by $ W $, we have
\begin{align}
J &= \frac{2d+1}{4\pi}v^{(0)T}v^{(0)}+\frac{1}{4}\sum_{l=1}^d W_l^T W_l +\frac{1}{2\pi d} \left(\sum_{\gamma=1}^d \!v^{(\gamma)T}v^{(\gamma)}
+\sum_{\alpha<\beta}\! v^{(\alpha,\beta)T}v^{(\alpha,\beta)}\!\right)+R,
\label{eigendecomp}
\end{align}
where the matrix $R$, which is defined as
\begin{eqnarray*}
R_{ij}=\frac{1}{2\pi}\sum_{n=1}^{\infty}\binom{2n}{n}\frac{(W^{(i)}\!\cdot \!W^{(j)})^{2n+2}}{2^{2n}(2n+1)(2n+2)(\norm{W^{(i)}}\norm{W^{(j)}})^{2n+1}},
\end{eqnarray*}
is positive-semidefinite.
\end{theorem}

In the above theorem, $W_l$ denotes the $l$-th row vector of $W$, and $v^{(0)}$
denotes the row vector whose $i$-th component is
$v^{(0)}_i=\|W^{(i)}\|/\sqrt{d}$, where $W^{(i)}\in \mathbb{R}^{d\times 1}$ denotes the $i$-th column vector of $W$.
We also define the row vector $v^{(\alpha,\beta)}$ whose $i$-th component is  
    $v^{(\alpha,\beta)}_i= \sqrt{d}W_{\alpha i}W_{\beta i}/\norm{W^{(i)}}$ ($1\leq \alpha\leq \beta\leq d$).
Further, define the row vector $v^{(\gamma)}$ by 
$v^{(\gamma)}= (v^{(\gamma,\gamma)}-v^{(0)})/\sqrt{2}$.
With high probability, the vectors $v^{(0)}$, ${\rm span}\{ W_l\}$, 
${\rm span}\{ v^{(\gamma)}\}$, and ${\rm span} \{ v^{(\alpha,\beta)}\}$
are approximately orthogonal, and the norms of $v^{(0)}$, $W_l$, $v^{(\gamma)}$, and $v^{(\alpha,\beta)}$
are close to $1$, assuming that each $W_{ij}$ is independently drawn from $N(0,1/m)$
and $m$ is sufficiently large.
Specifically, the following lemma holds (Lemma~3 in \cite{TakeishiNN}), where $\xi(d)=O(1/d^{1/2-\eta})$ for arbitrary $\eta\in(0,1/2)$. The definition of $\xi(d)$ is given in \cite{TakeishiNN}. 
\begin{lemma}[Takeishi et.al 2023]
\label{lem_orthnorm}
Let each $W_{ij}$ be realization of the independent random variable drawn from $N(0,1/m)$, 
let $C$ be a certain positive constant that does not depend on $m$ and $d$, $D=(d+1)(d+2)(d^2+3d+4)/8$, and $d>4$.
Then for all $\delta>0$, the following holds with the probability $1-CD/(\delta^2 m)$ at least:
\begin{enumerate}
  \item For the vectors $v^{(0)}$, $W_l$ $(1\leq l \leq d)$,
  $v^{(\alpha,\beta)}$ $(1\leq \alpha < \beta \leq d)$, and $v^{(\gamma)}$  $(1\leq \gamma \leq d)$, we have
  \begin{eqnarray*}
  |\|v^{(0)}\|^2-1| &\leq& \delta,\\
  |\|W_l\|^2-1| &\leq& \delta,\\
  |\|v^{(\alpha,\beta)}\|^2-1| &\leq& \delta+\xi(d),\\
  |\|v^{(\gamma)}\|^2-1| &\leq& \delta+\xi(d).
  \end{eqnarray*}
 
 \item Let $V_1=\{v^{(0)},W_l,v^{(\alpha,\beta)}|1\leq l \leq d,1\leq \alpha < \beta \leq d\}$,
       $V_2=\{v^{(\gamma)}|1\leq \gamma \leq d\}$, and $V=V_1\cup V_2$.
       We have for all different vectors $v,v'\in V$ ($v\notin V_2$ or $v'\notin V_2$)
       \begin{eqnarray*}
       |v\cdot v'|\leq \delta,
       \end{eqnarray*}
       and for all different vectors $v,v'\in V_2$
       \begin{eqnarray*}
       \left|v\cdot v' -\left(-\frac{1}{d-1}\right)\right|\leq \delta+\frac{\xi(d)}{d-1}.
       \end{eqnarray*}
\end{enumerate}
\end{lemma}

In fact, slightly modifying the normalization factors of $v^{(\alpha,\beta)}$ and $v^{(\gamma)}$, and refining the analysis, we can remove the terms involving $\xi(d)$ in this lemma. Furthermore, the reason why $v,v'\in V_2$ are nearly $-1/(d-1)$ in 2) is that we use $d$ vectors for the basis while  $\dim ({\rm span}\{ v^{(\gamma)}\})=d-1$ from the definition of $v^{(\gamma)}$. 
By introducing an new approximately orthonormal basis
$\{ \bar{v}^{(\gamma)} \}_{1\le \gamma \le d-1}$ of 
${\rm span}\{ v^{(\gamma)}\}$, we can also refine the evaluation of this part.

Now, we specifically describe the refinement of Lemma \ref{lem_orthnorm}. Define 
\begin{align}
\bar{v}^{(\alpha,\beta)}=\sqrt{1+\frac{2}{d}}\cdot{v}^{(\alpha,\beta)}\label{def_valpha}
\end{align}
and 
\begin{align}
\bar{v}^{(\gamma)}=\sqrt{1+\frac{2}{d}}\cdot\left(v^{(\gamma)}-\frac{1}{\sqrt{d}+1}v^{(d)}\right)\  (1\leq \gamma \leq d-1).\label{def_vgamma}
\end{align}
Then, we have the following lemma.
\begin{lemma}
\label{lem_orthnorm2}
Let each $W_{ij}$ be an independent random variable drawn from $N(0,1/p)$, 
let $\bar{C}$ be a certain positive constant that does not depend on $m$ and $d$, $\bar{D}=d(d+1)(d+2)(d+3)/8$, and $d>4$.
Let $\bar{V}=\{v^{(0)},W_l,\bar{v}^{(\alpha,\beta)},\bar{v}^{(\gamma)}|1\leq l \leq d,1\leq \alpha < \beta \leq d, 1\leq \gamma \leq d-1\}$.
Then for all $\delta>0$, the following holds with the probability $1-\bar{C}\bar{D}/(\delta^2 m)$ at least.
\begin{enumerate}
  \item For the vector $v\in \bar{V}$, we have
  $|\|v\|^2-1| \leq \delta$.
 \item For all different vectors $v,v'\in V$, we have $|v\cdot v'|\leq \delta$.
\end{enumerate}
\end{lemma}
Proof of this lemma is stated in Appendix \ref{append-2}.

By using the new basis $\bar{V}$, Theorem \ref{thm_decomp} is rewritten as follows.
\begin{corollary}
\label{cor_decomp}
Assume that
all the columns of matrix $W$ are non-zero vectors.
For the matrix $J=E[X^T X]$ and $m$-dimensional row vectors $v^{(0)} $, $ \{ W_l\}_{1\leq l\leq d} $, $ \{ \bar{v}^{(\gamma)}\}_{1\leq \gamma\leq d-1} $, $ \{ \bar{v}^{(\alpha,\beta)}\}_{1\leq \alpha< \beta\leq d} $ determined by $ W $, we have
\begin{align}
J &= \frac{2d+1}{4\pi}v^{(0)T}v^{(0)}+\frac{1}{4}\sum_{l=1}^d W_l^T W_l 
+\frac{1}{2\pi (d+2)} \left(\sum_{\gamma=1}^{d-1} \!\bar{v}^{(\gamma)T}\bar{v}^{(\gamma)}
+\sum_{\alpha<\beta}\! \bar{v}^{(\alpha,\beta)T}\bar{v}^{(\alpha,\beta)}\!\right)+R,
\label{eigendecomp_cor}
\end{align}
where the matrix $R$ is defined in Theorem \ref{thm_decomp}.
\end{corollary}

This decomposition is used for the analysis in the later part of this paper.
Let $\{ u^{(k)} \}_{1\leq k \leq D} $ denote the set of
$v^{(0)}$, $W_l$ ($1\le l \le d$), $\bar{v}^{(\gamma)}$ ($1\le \gamma \le d-1$),
and $v^{(\alpha,\beta)}$
($1\leq \alpha< \beta\leq d$), where $D=d(d+3)/2$.
For example, $u^{(1)}= v^{(0)}$, $u^{(2)}=W_1$, $u^{(d+2)}=\bar{v}^{(1)}$, etc.
Then, we can represent $J$ as
\begin{align}\label{Jappeigen}
    J &= \sum_{i=1}^D\tilde{\mu}_i u^{(i)T}u^{(i)}+R,
\end{align}
where 
\begin{align}\label{def_tillam}
\tilde{\mu}_{i} = 
\begin{cases}
(2d+1)/(4\pi) & \text{if $i=1$}\\
1/4  & \text{if $2\leq i\leq d+1$}\\
1/(2\pi (d+2))  & \text{if $d+2\leq i\leq D$}.
    \end{cases}
\end{align}


In this paper, we denote the spectral norm of a symmetric matrix $A$ as
\begin{align*}
\norm{A}_2=\max_{u:\|u\|=1} |uAu^T|,
\end{align*}
where $u$ is a row vector.
We denote the $D$-dimensional identity matrix as $I_D$.

We put the following assumption on $\{ u^{(k)} \}_{1\leq k \leq D} $,
which concerns how close it is to an orthonormal basis.
\begin{assumption}\label{assumption:gram}
Let $G=(g_{ij})$ denote the $D \times D$ matrix 
with $g_{ij}=u^{(i)}{u^{(j)}}^T$ 
(Gram matrix). 
Then for given $\epsilon_1>0$, $\norm{G-I_D}_2 \le  \epsilon_1$ 
holds.
\end{assumption}

When $m$ is large and $W_{ij}$ is independently drawn from $N(0,1/m)$,
the above assumption holds with a high probability, which will be  described in the next subsection.


\subsection{Risk Bound for Simple ReLU Networks}
In the following theorem, we show an upper bound on the right-hand side of \eqref{eqn_riskbound} for the simple ReLU network model.
\begin{theorem}\label{thm_redbound2}
Let $\alpha>1$ be a hyper parameter of the MDL estimator.
We can design an MDL estimator for the simple ReLU network model $\mathcal{M}=\{p_v(y|x) : \norm{v} \leq 1\}$
, which satisfies the following.
Under Assumption \ref{assumption:gram} for $0<\epsilon_1<1$, 
for each $n$, the risk is bounded as
\begin{align}
\mathbb{E} \left[d_{\lambda}(p^*, \ddot{p})\right]
&\leq \frac{\alpha D(\log n+1)}{2n}+\frac{\alpha}{n}\sum_{i\leq D}
\left(\log c_i +\frac{1}{c_i\sqrt{n}}\right)
+\frac{2(1+\epsilon_1)(\beta +\epsilon'_1 ) \mathrm{tr}(J)}{\sigma^2},
\label{eqn_riskbound_relu}
\end{align}
where $\beta=1-\sum_{i=1}^D \tilde{\mu}_i/\mathrm{tr}(J)$, $\epsilon'_1=\epsilon_1/(1-\epsilon_1)$, 
and $c_i = (1+\epsilon_1) \sqrt{\tilde{\mu}_i/(\alpha\sigma^2)}$.
\end{theorem}

The right-hand side of equation \eqref{eqn_riskbound_relu} does not depend on $m$ since the following lemma holds.
\begin{lemma}
\label{lem_assump}
Let each $W_{ij}$ be an independent random variable drawn from $N(0,1/p)$, 
let $\bar{C}$ be a certain positive constant that does not depend on $m$ and $d$, $\bar{D}=d(d+1)(d+2)(d+3)/8$, and $d>4$.
Then for all $\epsilon_1>0$, the following holds with the probability $1-\bar{C}\bar{D}^3/(\epsilon_1^2 m)$ at least.
\begin{enumerate}
  \item Assumption \ref{assumption:gram} holds.
  \item We have
  \begin{align}
  \mathrm{tr}(J)\leq \frac{d}{2}\left(1+\frac{\epsilon_1}{\bar{D}}\right).
  \end{align}
  \item We have
  \begin{align}
  \beta\leq 0.023 +\frac{\epsilon_1}{\bar{D}}.
  \end{align}
\end{enumerate}
\end{lemma}

{\it Proof of Lemma \ref{lem_assump}:} We denote the Frobenius norm of a matrix $A$ as 
\[\norm{A}_F=\sqrt{\sum_{i,j}|A_{ij}|^2}.\]
According to Lemma \ref{lem_orthnorm2}, the following holds with the probability $1-\bar{C}\bar{D}/(\delta^2 m)$ at least.
\begin{enumerate}
  \item We have
\begin{align}
\norm{G-I_D}_2 \le \norm{G-I_D}_F \le \sqrt{\sum_{i,j}\delta^2}=D\delta.
\end{align}

  \item The trace of the matrix $J$ is
  \begin{align}
  \mathrm{tr}(J)= \sum_{i=1}^m J_{ii}= \frac{1}{2}\sum_{i=1}^m\|W^{(i)}\|^2 =\frac{1}{2}\sum_{l=1}^d \|W_l\|^2.
  \end{align}
  From Lemma \ref{lem_orthnorm2}, we have for all $1\leq l \leq d$, 
  $
  \left|\|W_l\|^2-1\right| \leq \delta.
  $
  Then, $\mathrm{tr}(J)$ satisfies
  \begin{align}
  \mathrm{tr}(J)\leq \frac{d}{2}(1+\delta).
  \end{align}

\item From the definition of $\tilde{\mu}_i$ \eqref{def_tillam}, we have
\begin{align}
\sum_{i=1}^D \tilde{\mu}_i&=\frac{2d+1}{4\pi}+\frac{1}{4}\cdot d+\frac{1}{2\pi (d+2)}\cdot\left(\frac{d(d-1)}{2}+d-1\right)\\
&=\frac{2d+1}{4\pi}+\frac{d}{4}+\frac{d-1}{4\pi}\\
&=
\frac{3+\pi}{2\pi}\cdot \frac{d}{2}\geq 0.977\cdot \frac{d}{2},
\end{align}
By using 2), we have
\begin{align}
\beta = 1-\frac{\sum_{i=1}^D \tilde{\mu}_i}{\mathrm{tr}(J)}\leq 1 - \frac{0.977}{1+\delta}\leq 0.023+\delta.
\end{align}

\end{enumerate}
By substituting $\delta=\epsilon_1/\bar{D}$, we obtain the lemma.

\subsection{Design of Two-Stage Code}
Here, we will design our two-stage code for the two-layer ReLU networks
making use of the approximate spectral decomposition of $J$.
Setting $D=d(d+3)/2$, let $V_D$ denote ${\rm span}\{u^{(1)},u^{(2)},\ldots,u^{(D)} \}$, where the vectors are defined in Section \ref{SecFIM}. 
Let $\{ u^{(D+1)} ,\ldots,u^{(m)} \}$ be
an orthonormal basis of $V_D^{\perp}$.
Then, let $\{ \tilde{u}^{(i)} \}_{1\le i \le m}$
denote the dual basis of $\{ u^{(i)} \}_{1\le i \le m}$,
that is, $u^{(i)}\tilde{u}^{(j)T} = \delta_{ij}$ holds.
Here, $\delta_{ij}$ denotes the Kronecker's delta.
Using the dual basis,
define the $m\times m$ matrix $K$ as
\begin{align}
    K = (\tilde{u}^{(1)T},\tilde{u}^{(2)T},\ldots,\tilde{u}^{(m)T})^T.
\end{align}
Note that 
\begin{align}
    K^{-1}
=(u^{(1)T},u^{(2)T},\ldots,u^{(D) T},u^{(D+1) T},\ldots, u^{(m) T})
\end{align}
holds.
Then, 
we introduce a new parameter $\theta$ by $\theta= vK^{-1}$

Since $u^{(i)}$ with $i > D$ is orthogonal to any $u^{(i)}$ with $i \le D$,
$(KK^T)^{-1}=(K^{-1})^TK^{-1}$ is block-diagonal with block sizes $D$ and $m-D$,
where one block is $G$ and the other is the unit matrix of order $m-D$.

Hence, under Assumption~\ref{assumption:gram},
\begin{align}
\norm{(KK^T)^{-1}-I_m}_2=\norm{G-I_D}_2 \le \epsilon_1
\label{asm1-2}
\end{align}
and
\begin{align}
\norm{u}^2 = v K^{-1} (K^{-1})^T v^T &\le 
\norm{K^{-1} (K^{-1})^T}_2\norm{v}^2\\
&=\norm{(KK^T)^{-1}}_2\norm{v}^2
\end{align}
 hold.
From \eqref{asm1-2}, we have 
\begin{align}
\norm{(KK^T)^{-1}}_2 \le 1+\epsilon_1.
\end{align}
Therefore, $\Theta = \{ \theta= vK^{-1} | \|v\| \le 1 \}$ is included
in the sphere centered at the origin with radius $\sqrt{1+\epsilon_1}$.

The quantization strategy for the parameter space $\Theta$ is essentially the same as in Section \ref{linearmodel}, but the differences are described below. First, the $\tilde{\mu}_i$ used in the definition of the quantization width $\Delta_i=2\sqrt{\alpha\sigma^2/(n\tilde{\mu}_i)}$ is not the exact eigenvalue, but the approximate eigenvalue discussed in this section. Additionally, the range of quantization to be considered in the $\theta_i$ direction should be $[-\sqrt{1+\epsilon_1}, \sqrt{1+\epsilon_1}]$, thus the number of quantization points is $q_i=\left\lceil2\sqrt{1+\epsilon_1}/\Delta_i\right\rceil$. By defining $c_i$ used in the evaluation of $L_i$ as $c_i = \sqrt{(1+\epsilon_1)\tilde{\mu}_i/(\alpha\sigma^2)}$, \eqref{param_length} holds similarly.

\subsection{Proof of Theorem \ref{thm_redbound2}}
The proof is carried out in the same manner as Theorem \ref{thm_redbound}, but due to the difference in the definition of $K$, the evaluation of \eqref{quadbound} is performed as follows.
By \eqref{Jappeigen} and $u^{(i)} \tilde{u}^{(j)T} = \delta_{ij}$, we have
\begin{align}\label{fortrace}
    (\theta-\theta^*) K J K^T (\theta-\theta^*)^T = \sum_{i=1}^D \tilde{\mu}_i (\theta_i-\theta^*_i)^2 + (\theta-\theta^*) K R K^T (\theta-\theta^*)^T.
\end{align}

The matrix $KK^T$ is block-diagonal with two blocks whose sizes are $D$ and $m-D$ respectively.
Here, since one block is $G^{-1}$ and the other is the unit matrix of order $m-D$, 
we have $\norm{KK^T}_2=\norm{G^{-1}}_2$. 
Let $\bar{u}$ denote the eigenvector of $G^{-1}$ corresponding the largest eigenvalue. Then, we have for all $u:\|u\|=1$
\begin{align*}
&|uG^{-1}u^T|\leq |\bar{u}G^{-1}\bar{u}^T|=\frac{1}{|\bar{u}G\bar{u}^T|}\leq \frac{1}{1-\epsilon_1},
\end{align*}
where the last inequality holds from Assumption \ref{assumption:gram}.
Thus, letting $\epsilon'_1=\epsilon_1/(1-\epsilon_1)$, we have $\norm{KK^T}_2 \le 1+\epsilon'_1$. This fact yields
\begin{align}
    \mathrm{tr}(KJK^T) = \mathrm{tr}(K^TKJ) \le (1+\epsilon'_1)\mathrm{tr}(J)
    .
\end{align}
On the other hand, we have
\begin{align}
    \mathrm{tr}(KJK^T) =\sum_{i=1}^D \tilde{\mu}_i 
    + \mathrm{tr}(KRK^T).
\end{align}
Hence, we have
\begin{align}
    \mathrm{tr}(KRK^T) \le
     (1+\epsilon'_1)\mathrm{tr}(J) - \sum_{i=1}^D \tilde{\mu}_i=(\beta+\epsilon'_1)\mathrm{tr}(J).
\end{align}

Noting that $\Theta \subset \{ \theta | \norm{\theta}^2 \le 1+\epsilon_1 \}$, we have
for all $\theta, \theta^* \in \Theta$
\begin{align}\label{trKRK}
    (\theta-\theta^*) KRK^T (\theta-\theta^*)^T  
    \leq \norm{\theta-\theta^*}^2 \mathrm{tr}(KRK^T)
    \leq
    4(1+\epsilon_1)(\beta +\epsilon'_1 )\mathrm{tr}(J).
\end{align}
Therefore, in this case, \eqref{quadbound} is evaluated as
\begin{align}
    (\theta-\theta^*) K J K^T (\theta-\theta^*)^T \leq \sum_{i=1}^D \tilde{\mu}_i (\theta_i-\theta^*_i)^2 + 4(1+\epsilon_1)(\beta +\epsilon'_1 )\mathrm{tr}(J).
\end{align}
From this, the theorem is derived.

\subsection{Remarks on our Risk Bound}
From Theorem \ref{thm_redbound2}, the risk bound of the MDL estimator is given in the right-hand side of \eqref{eqn_riskbound_relu}.
The first term takes the form $O(D\log n/n)$, which correspond the risk bound of the traditional MDL estimator for a parameter space of dimension $D$.
The third term, $2(1+\epsilon_1)(\beta +\epsilon'_1) \mathrm{tr}(J)/\sigma^2$,  denotes the error in the direction $u_{D+1},\ldots,u_m$, where we have given up on the encoding.  
Note that when $m$ is sufficiently large, the overall upper bound does not depend on $m$.

Although $D$ is defined as $d(d+3)/2$ in this paper, $\beta$ can be further reduced if $D$ is made larger.
To do so, we need to examine the $d(d+3)/2+1$-th and subsequent terms in the approximate eigenvalue decomposition of $J$ in \eqref{Jappeigen}. 

As $L(u)$ is constant in this case, our MDL estimator performs maximum likelihood estimation over the quantized parameter space, focusing only on the primary $D$ eigenvector directions.
The behavior of our estimator shares noticeable similarities with that of Gradient Descent (GD), widely used in the training of NNs.
In GD, as iterations increase, learning progressively advances from the direction of eigenvectors corresponding to larger eigenvalues of the FIM (refer to Section 6 in \cite{TakeishiNN}). 
If we implement 'early stopping' after the completion of learning in $D$ directions, this learning process resembles our MDL estimator.
Our current analysis is limited to the learning of the final layer of a two-layer NNs, but this could potentially offer clues to understand why learning with GD is successful in deep NNs.

\subsection{Analysis of Model Misspecification}
We analyze the case where the true distribution is not necessarily included in the model $\mathcal{M}$. 
Assume that for a function $f^*:\mathbb{R}^d\rightarrow \mathbb{R}$, $y = f^*(x) + \epsilon$.
In this case, the true distribution is $p^*(y|x) = N(y|f^*(x), \sigma^2)$.
Also in this case, Theorem \ref{thm_riskbound} holds.
Then, by rewriting the right hand side of \eqref{eqn_riskbound}, we have 
\begin{align}
\mathbb{E} \left[d_{\tilde{\mu}}(p^*, \ddot{p})\right]
&\leq \frac{1}{n} 
\mathbb{E} \left[\log \frac{p_{\tilde{v}}(y^n|x^n)}{{p}_{\ddot{v}}(y^n|x^n)e^{-\alpha L(\ddot{v})}}+\log \frac{p^*(y^n|x^n)}{{p}_{\tilde{v}}(y^n|x^n)}\right].\\
&=\frac{1}{n} 
\mathbb{E} \left[\log \frac{p_{\tilde{v}}(y^n|x^n)}{{p}_{\ddot{v}}(y^n|x^n)e^{-\alpha L(\ddot{v})}}\right]+D_{KL}(p^*||p_{\tilde{v}}).
\label{mis_red}
\end{align}
Here, we defined 
\[\tilde{v}=
\arg\min_{v \in V} D_{KL}(p^*||p_{v})=
\arg \min_{v \in V} \mathbb{E}
\left[-\log {p}_{v}(y|x)\right],\] where this expectation is taken for $(x,y) \sim p^*(y|x)p(x)$, and $D_{KL}(p^*||p_{\tilde{v}})$ denotes the KL divergence between $p^*$ and $p_{\tilde{v}}$.
Further, assume that $\tilde{v}$ is not on the boundary of $V = \{v \in \mathbb{R}^m : \norm{v} \leq 1\}$, which is equivalent to saying that $\norm{\tilde{v}} < 1$. Then, by the definition of $\tilde{v}$, we have
\begin{align}
\left.\nabla\mathbb{E} \left[-\log {p}_{v}(y|x)\right]\right|_{v=\tilde{v}}
=-\mathbb{E} \left[\left.\nabla\log {p}_{v}(y|x)\right|_{v=\tilde{v}}\right]=0.\label{tildev}
\end{align}
Under this assumption, we can similarly analyze the first term of \eqref{mis_red} as the proof of Theorem \ref{thm_redbound2}, and we have the following theorem.
\begin{theorem}\label{thm_redbound_mis}
Let $\alpha>1$ be a hyper parameter of the MDL estimator.
We can design an MDL estimator for the simple ReLU network model $\mathcal{M}=\{p_v(y|x) : \norm{v} \leq 1\}$, which satisfies the following.
Assume that $\norm{\tilde{v}}<1$. Under Assumption \ref{assumption:gram} for $0<\epsilon_1<1$, for each $n$, the risk is bounded as
\begin{align}
\mathbb{E} \left[d_{\lambda}(p^*, \ddot{p})\right]
&\leq \frac{\alpha D(\log n+1)}{2n}+\frac{\alpha}{n}\sum_{i\leq D}
\left(\log c_i +\frac{1}{c_i\sqrt{n}}\right)
+\frac{2(1+\epsilon_1)(\beta +\epsilon'_1 ) \mathrm{tr}(J)}{\sigma^2}+D_{KL}(p^*||p_{\tilde{v}}),
\label{eqn_riskbound_relu_mis}
\end{align}
where $\beta=1-\sum_{i=1}^D \tilde{\mu}_i/\mathrm{tr}(J)$, $\epsilon'_1=\epsilon_1/(1-\epsilon_1)$, 
and $c_i = (1+\epsilon_1)\sqrt{\tilde{\mu}_i/(\alpha\sigma^2)}$.
\end{theorem}

{\it Proof:} We evaluate the expectation 
\begin{align}
\mathbb{E} \left[\log \frac{p_{\tilde{v}}(y^n|x^n)}{p_{\ddot{v}}(y^n|x^n)e^{-\alpha L(\ddot{v})}}\right]
=\mathbb{E} \left[\log \frac{\bar{p}_{\tilde{\theta}}(y^n|x^n)}{\bar{p}_{\ddot{\theta}}(y^n|x^n)e^{-\alpha L(\ddot{\theta})}}\right]
\end{align}
in the right hand side of \eqref{mis_red} under the variable transformation $\theta=vK^{-1}$. Then, we have for an arbitrary $\theta$, 
\begin{align}
    \log\frac{\bar{p}_{\tilde{\theta}}(y^n|x^n)}{\bar{p}_\theta(y^n|x^n)}
    =&
    - \left.\nabla \log \bar{p}_\theta(y^n|x^n)\right|_{\theta=\tilde{\theta}}(\theta-\tilde{\theta})^T
    +
    \frac{n(\theta-\tilde{\theta})K\hat{J}K^T(\theta-\tilde{\theta})^T}{2\sigma^2}.
\end{align}
Noting that 
$\mathbb{E} \left[\left.\nabla\log \bar{p}_\theta(y|x)\right|_{\theta=\tilde{\theta}}\right]=0$
from \eqref{tildev}, we have
\begin{align}
  \mathbb{E}\Bigl[
  \log\frac{\bar{p}_{\tilde{\theta}}(y^n|x^n)}{\bar{p}_\theta(y^n|x^n)}
\Bigr]
    =
    \frac{n(\theta-\tilde{\theta})KJK^T(\theta-\tilde{\theta})^T}{2\sigma^2}.
\end{align}
Recalling the definition of $\ddot{\theta}$, we have for an arbitrary $\theta \in \ddot{\Theta}_n$
\begin{align}
  \mathbb{E}\Bigl[
  \log\frac{\bar{p}_{\tilde{\theta}}(y^n|x^n)}{\bar{p}_{\ddot{\theta}}(y^n|x^n)e^{-\alpha L(\ddot{\theta}))}}
\Bigr]\le
    \frac{n(\theta-\tilde{\theta})KJK^T(\theta-\tilde{\theta})^T}{2\sigma^2}+\alpha L(\theta).
\end{align}
The discussion after \eqref{expRn} in the proof of Theorem \ref{thm_redbound} also holds if $\theta^*$ is replaced  to $\tilde{\theta}$.
By using it, we obtain the theorem.




\section*{Acknowledgments}
The authors thank Professor Andrew Barron, Professor Hiroshi Nagaoka, Professor Noboru Murata, and Professor Kazushi Mimura  for their helpful comments.



%

\appendices

\section{Proof of Theorem \ref{thm_riskbound}}
\label{append-1}
Since $d_{\lambda}(p_1,p_2)$ is increasing in $\lambda \in (0,1)$,
the proof for the case that
$\lambda = 1-\alpha^{-1}$ is sufficient.
Hence we assume it.
We have
\begin{align}\label{main_manipu_BC}
    &\E\Bigl( d_{\lambda}(p^*, \ddot{p}) -\frac{1}{n} \log\frac{p^*(y^n|x^n)}{q(y^n|x^n)} \Bigr) \\
    = &
    \E\Bigl( d_{\lambda}(p^*, \ddot{p}) -\frac{1}{n}\log\frac{p^*(y^n|x^n)}{\ddot{p}(y^n|x^n)} 
    -\frac{\alpha L(\ddot{v})}{n}\Bigr)
    \\
    =&
    \frac{\alpha}{n}
    \E\Bigl( \frac{n d_{\lambda}(p^*, \ddot{p})}{\alpha} -\frac{1}{\alpha}\log\frac{p^*(y^n|x^n)}{\ddot{p}(y^n|x^n)} 
    - L(\ddot{v})\Bigr)
        \\
    =&
    \frac{\alpha}{n}
    \E\log \exp \Bigl( \frac{n d_{\lambda}(p^*, \ddot{p})}{\alpha} -\frac{1}{\alpha}\log\frac{p^*(y^n|x^n)}{\ddot{p}(y^n|x^n)} 
    - L(\ddot{v})\Bigr)
            \\
    \le &
    \frac{\alpha}{n}
    \log \E \exp \Bigl( \frac{n d_{\lambda}(p^*, \ddot{p})}{\alpha} -\frac{1}{\alpha}\log\frac{p^*(y^n|x^n)}{\ddot{p}(y^n|x^n)} 
    - L(\ddot{v})\Bigr)
    \\
        \le
    &
     \frac{\alpha}{n}
    \log \sum_{\bar{v}\in \ddot{V}_n} \E \exp \Bigl( \frac{n d_{\lambda}(p^*, p_{\bar{v}})}{\alpha} 
    -\frac{1}{\alpha}\log\frac{p^*(y^n|x^n)}{p_{\bar{v}}(y^n|x^n)} 
    -L(\bar{v})\Bigr).
\end{align}
The first inequality follows from Jensen's inequality.
The summation in the last line is
\begin{align}
  \sum_{\bar{v}\in \ddot{V}_n} \E \exp \Bigl( \frac{n d_{\lambda}(p^*, p_{\bar{v}})}{\alpha} 
    -\frac{1}{\alpha}\log\frac{p^*(y^n|x^n)}{p_{\bar{v}}(y^n|x^n)} 
    -L(\bar{v})\Bigr)
    =
 \sum_{\bar{v}\in \ddot{V}_n}  \exp \Bigl( \frac{n d_{\lambda}(p^*, p_{\bar{v}})}{\alpha} 
    \Bigr)  e^{-L(\bar{v})}
    \E\Bigl(\frac{p_{\bar{v}}(y^n|x^n)}{p^*(y^n|x^n)} \Bigr)^{1/\alpha},
\end{align}
where
\begin{align}
    \E\Bigl(\frac{p_{\bar{v}}(y^n|x^n)}{p^*(y^n|x^n)} \Bigr)^{1/\alpha}
=
\E\Bigl(\prod_t\frac{p_{\bar{v}}(y_t|x_t)}{p^*(y_t|x_t)} \Bigr)^{1/\alpha}
=\prod_t
\E\Bigl(\frac{p_{\bar{v}}(y_t|x_t)}{p^*(y_t|x_t)} \Bigr)^{1-\lambda}
=\left\{\E\Bigl(\frac{p_{\bar{v}}(y|x)}{p^*(y|x)} \Bigr)^{1-\lambda}\right\}^n.
\end{align}
Further, recalling the definition of R\'enyi divergence
\begin{align}
    \exp \Bigl( \frac{n d_{\lambda}(p^*, p_{\bar{v}})}{\alpha} 
    \Bigr)
    & =
    \exp (n (1-\lambda)d_{\lambda}(p^*, p_{\bar{v}}))\\
    & =
    \exp\Bigl(n (1-\lambda) \Bigl(-\frac{1}{1-\lambda} \log \E \Bigl(\frac{p_{\bar{v}}(y|x)}{p^*(y|x)}\Bigr)^{1-\lambda}\Bigr)\Bigr)\\
    &=\left\{\E\Bigl(\frac{p_{\bar{v}}(y|x)}{p^*(y|x)} \Bigr)^{1-\lambda}\right\}^{-n}.
\end{align}
Hence
\begin{align}\label{proof_of_BC}
  \sum_{\bar{v}\in \ddot{V}_n} \E \exp \Bigl( \frac{n d_{\lambda}(p^*, p_{\bar{v}})}{\alpha} 
    -\frac{1}{\alpha}\log\frac{p^*(y^n|x^n)}{p_{\bar{v}}(y^n|x^n)} 
    -L(\bar{v})\Bigr)
=\sum_{\bar{v}\in \ddot{V}_n}e^{-L(\bar{v})}\le 1,
\end{align}
which implies
\begin{align}
       \E\Bigl( d_{\lambda}(p^*, \ddot{p}) -\frac{1}{n} \log\frac{p^*(y^n|x^n)}{q(y^n|x^n)} \Bigr) 
    \le 0.
\end{align}
This completes the proof of the theorem.

\section{Proof of Lemma \ref{lem_orthnorm2}}
\label{append-2}
To prove Lemma \ref{lem_orthnorm2}, we use the following lemma.
\begin{lemma}
\label{lem_exp_var2}
Let each $W_{ij}$ be an independent random variable drawn from $N(0,1/m)$, 
let $\bar{C}$ be a certain positive constant that does not depend on $m$ and $d$, and $d>4$.
Let $\bar{V}=\{v^{(0)},W_l,\bar{v}^{(\alpha,\beta)},\bar{v}^{(\gamma)}|1\leq l \leq d,1\leq \alpha < \beta \leq d, 1\leq \gamma \leq d-1\}$.
Then, the following holds, where the expectation and the variance are taken with respect to $W$.
\begin{enumerate}
  \item For the vector $v\in \bar{V}$, we have
  $\mathbb{E}[\|v\|^2]=1$.
 \item For all different vectors $v,v'\in \bar{V}$, we have $\mathbb{E}[v\cdot v']=0$.
 \item For all vectors $v,v'\in \bar{V}$, the variance of the inner product is bounded as $Var[v\cdot v']\leq \bar{C}/m$.
\end{enumerate}
\end{lemma}

Proof of Lemma \ref{lem_exp_var2} is stated in Appendix \ref{pf_lem_exp_var2}. Note that Lemma \ref{lem_exp_var2} is a refined version of the following lemma, which is used for the proof of Lemma \ref{lem_orthnorm} \cite{TakeishiNN}.
\begin{lemma}[Takeishi et.al 2023]
\label{lem_exp_var}
Let each $W_{ij}$ be an independent random variable drawn from $N(0,1/m)$, 
let $C$ be a certain positive constant that does not depend on $m$ and $d$, and $d>4$. Then, the following holds, where the expectation and the variance are taken with respect to $W$.
\begin{enumerate}
  \item For the vectors $v^{(0)}$, $W_l$ $(1\leq l \leq d)$,
  $v^{(\alpha,\beta)}$ $(1\leq \alpha < \beta \leq d)$, and $v^{(\gamma)}$  $(1\leq \gamma \leq d)$, we have
  \begin{eqnarray*}
  \mathbb{E}[\|v^{(0)}\|^2] &=& 1,\\
  \mathbb{E}[\|W_l\|^2] &=& 1,\\
  |\mathbb{E}[\|v^{(\alpha,\beta)}\|^2]-1| &\leq& \xi(d),\\
  |\mathbb{E}[\|v^{(\gamma)}\|^2]-1| &\leq& \xi(d).
  \end{eqnarray*}
 
 \item Let $V_1=\{v^{(0)},W_l,v^{(\alpha,\beta)}|1\leq l \leq d,1\leq \alpha < \beta \leq d\}$,
       $V_2=\{v^{(\gamma)}|1\leq \gamma \leq d\}$, and $V=V_1\cup V_2$.
       We have for all different vectors $v,v'\in V$ ($v\notin V_2$ or $v'\notin V_2$)
       \begin{eqnarray*}
       \mathbb{E}[v\cdot v']=0,
       \end{eqnarray*}
       and for all different vectors $v,v'\in V_2$
       \begin{eqnarray*}
       \left|\mathbb{E}[v\cdot v'] -\left(-\frac{1}{d-1}\right)\right|\leq \frac{\xi(d)}{d-1}.
       \end{eqnarray*}
 \item For all vectors $v,v'\in V$, the variance of the inner product is bounded as
       \begin{eqnarray*}
       Var[v\cdot v']\leq \frac{C}{m}.
       \end{eqnarray*}
\end{enumerate}
\end{lemma}

Now, we describe the proof of Lemma \ref{lem_orthnorm2}.
Define the set of random variables $\bar{V}_{pair}=\{v\cdot v'|v,v' \in \bar{V} \}$.
Since the cardinality of $\bar{V}$ is $d(d+3)/2$, that of $\bar{V}_{pair}$ is 
\begin{align}
\frac{1}{2}\cdot\frac{d(d+3)}{2}\cdot\left(\frac{d(d+3)}{2}+1\right)=\frac{d(d+1)(d+2)(d+3)}{8}=\bar{D}.
\end{align}
From Lemma \ref{lem_exp_var2}, it follows that $Var(T)\leq \bar{C}/m$ for all $T \in \bar{V}_{pair}$, where $\bar{C}$ is a certain positive constant.
Thus, from the Chebyshev's inequality, we have for all $T \in \bar{V}_{pair}$
\begin{eqnarray*}
\Pr(|T-\mathbb{E}(T)|\geq \delta)\leq \frac{\bar{C}}{\delta^2 m}.
\end{eqnarray*}
Further, using the union bound for the above, we have 
\begin{eqnarray}
&&\Pr(\exists T \in \bar{V}_{pair},\ |T-\mathbb{E}(T)|\geq \delta)\leq \frac{\bar{C}\bar{D}}{\delta^2 m}\nonumber\\
&\Leftrightarrow&\Pr(\forall T \in \bar{V}_{pair},\ |T-\mathbb{E}(T)|\leq \delta)\geq 1-\frac{\bar{C}\bar{D}}{\delta^2 m}.\label{event}
\end{eqnarray}
From Lemma \ref{lem_exp_var2}, $\mathbb{E}(T)=1$ when $T\in\{\|v\|^2 : v \in \bar{V}\}$, and $\mathbb{E}(T)=0$ otherwise. This completes the proof of lemma.

\section{Proof of Lemma \ref{lem_exp_var2}}
\label{pf_lem_exp_var2}
It is sufficient that we prove the different parts of the lemma from Lemma \ref{lem_exp_var}.
We let the $d$-dimensional random vector $Z$ each element of which  follows a standard normal distribution, and $Z_\alpha$ denote the $\alpha$-th component of $Z$.
Now, we will prove 1), 2), and 3) in the lemma.
\begin{enumerate}
\item 
In fact, we can explicitly calculate the expectations $\mathbb{E}[\|v^{(\alpha,\beta)}\|^2]$ and $\mathbb{E}[\|v^{(\gamma)}\|^2]$, while their calculation was previously considered difficult in \cite{TakeishiNN}.
First, we have
\begin{align}
\mathbb{E}[\|v^{(\alpha,\beta)}\|^2]=\sum_{i=1}^p \mathbb{E}\left[\frac{d W_{\alpha i}^2 W_{\beta i}^2}{\|W^{(i)}\|^2}\right]
=d\mathbb{E}\left[\frac{Z_{\alpha}^2 Z_{\beta}^2}{\|Z\|^2}\right].
\end{align}
Letting $U=Z/\|Z\|$ and $\tilde{X}=\|Z\|^2$, we have
\begin{align}
\mathbb{E}\left[\frac{Z_{\alpha}^2 Z_{\beta}^2}{\|Z\|^2}\right]
=\mathbb{E}[U_{\alpha}^2 U_{\beta}^2\tilde{X}]=\mathbb{E}[U_{\alpha}^2 U_{\beta}^2]\cdot\mathbb{E}[\tilde{X}],
\end{align}
where we used the fact $U$ and $\tilde{X}$ are independent since $U$ follows a spherical uniform distribution on a unit sphere.
Now, we calculate $\mathbb{E}[U_{\alpha}^2 U_{\beta}^2]$. Since 
\begin{align}
\mathbb{E}[U_{\alpha}^2 U_{\beta}^2]\cdot\mathbb{E}[\tilde{X}^2]=\mathbb{E}[U_{\alpha}^2 U_{\beta}^2\tilde{X}^2]
=\mathbb{E}\left[Z_{\alpha}^2 Z_{\beta}^2\right]=1,
\end{align}
we have 
\begin{align}
\mathbb{E}[U_{\alpha}^2 U_{\beta}^2]=\frac{1}{\mathbb{E}[\tilde{X}^2]}.
\end{align}
Noting that $\tilde{X}$ is subject to chi-squared distribution, we have
\begin{align}
\mathbb{E}[\tilde{X}^2]=(\mathbb{E}[\tilde{X}])^2+Var[\tilde{X}]=d^2+2d.
\end{align}
Thus, we have
\begin{align}
\mathbb{E}[\|v^{(\alpha,\beta)}\|^2]=
 \frac{d\mathbb{E}[\tilde{X}]}{\mathbb{E}[\tilde{X}^2]}=\frac{d^2}{d^2+2d}=\frac{d}{d+2}.
\end{align}
Recalling the definition of $\bar{v}^{(\alpha,\beta)}$ \eqref{def_valpha}, we have
\begin{align}
\mathbb{E}[\|\bar{v}^{(\alpha,\beta)}\|^2]=\frac{d+2}{d}
\mathbb{E}[\|v^{(\alpha,\beta)}\|^2]
=1.
\end{align}

Secondly, we have
\begin{align}
\mathbb{E}[\|v^{(\gamma,\gamma)}\|^2]=\sum_{i=1}^p \mathbb{E}\left[\frac{d W_{\gamma i}^4}{\|W^{(i)}\|^2}\right]
=d\mathbb{E}\left[\frac{Z_{\gamma}^4}{\|Z\|^2}\right].
\end{align}
Similarly, we have
\begin{align}
\mathbb{E}\left[\frac{Z_{\gamma}^4}{\|Z\|^2}\right]
=\mathbb{E}[U_{\gamma}^4\tilde{X}]=\mathbb{E}[U_{\gamma}^4]\cdot\mathbb{E}[\tilde{X}].
\end{align}
Since 
\begin{align}
\mathbb{E}[U_{\gamma}^4]\cdot\mathbb{E}[\tilde{X}^2]=\mathbb{E}[U_{\gamma}^4\tilde{X}^2]
=\mathbb{E}\left[Z_{\gamma}^4\right]=3,
\end{align}
we have 
\begin{align}
\mathbb{E}[U_{\gamma}^4]=\frac{3}{\mathbb{E}[\tilde{X}^2]}=\frac{3}{d^2+2d}.
\end{align}
Noting that $\tilde{X}$ is subject to chi-squared distribution, we have
\begin{align}
\mathbb{E}[\tilde{X}^2]=(\mathbb{E}[\tilde{X}])^2+Var[\tilde{X}]=d^2+2d.
\end{align}
Thus, we have
\begin{align}
\mathbb{E}[\|v^{(\gamma,\gamma)}\|^2]=\frac{3d}{d+2}.
\end{align}
Recalling the definition of ${v}^{(\gamma)}$, we have
\begin{align}
\mathbb{E}[\|v^{(\gamma)}\|^2]
&=\frac{1}{2}\mathbb{E}[\|v^{(\gamma,\gamma)}-v^{(0)}\|^2]\nonumber\\
&=\frac{1}{2}\mathbb{E}[\|v^{(\gamma,\gamma)}\|^2-2v^{(\gamma,\gamma)}\cdot v^{(0)} +\|v^{(0)}\|^2]\nonumber\\
&=\frac{1}{2}\left(\frac{3d}{d+2}-2+1\right)\\
&=\frac{d-1}{d+2},
\label{norm_vgamma}
\end{align}
where we have used 
\begin{align}
\mathbb{E}[v^{(\gamma,\gamma)}\cdot v^{(0)}]=\sum_{i=1}^p \mathbb{E}\left[\frac{\sqrt{d}W_{\gamma i}^2}{\|W^{(i)}\|}\cdot 
\frac{\|W^{(i)}\|}{\sqrt{d}} \right]
=\sum_{i=1}^p \mathbb{E}[W_{\gamma i}^2]=1
\end{align} 
and $\mathbb{E}[\|v^{(0)}\|^2]=1$.
Furthermore, we have
\begin{align}
\mathbb{E}\left[\left\|\sum_{\gamma=1}^d v^{(\gamma)}\right\|^2\right]
&=\mathbb{E}\left[\sum_{\gamma=1}^d \|v^{(\gamma)}\|^2 + \sum_{\gamma\neq\gamma'}^d v^{(\gamma)}\cdot v^{(\gamma')} \right]\\
&=d\mathbb{E}\left[\|v^{(\gamma)}\|^2\right]+d(d-1)\mathbb{E}\left[v^{(\gamma)}\cdot v^{(\gamma')} \right].
\end{align}
Noting that $\sum_{\gamma=1}^d v^{(\gamma)}= 0$, we have $\gamma\neq\gamma'$
\begin{align}
\mathbb{E}\left[v^{(\gamma)}\cdot v^{(\gamma')}\right]=-\frac{1}{d-1}\mathbb{E}\left[\|v^{(\gamma)}\|^2\right].
\label{prod_vgamma}
\end{align}
Then, we have for $1\leq \gamma \leq d-1$
\begin{align}
\mathbb{E}\left[\left\|v^{(\gamma)}-\frac{1}{\sqrt{d}+1}v^{(d)}\right\|^2\right]
&=\mathbb{E}\left[\|v^{(\gamma)}\|^2+\frac{1}{(\sqrt{d}+1)^2}\|v^{(d)}\|^2-\frac{2v^{(\gamma)}\cdot v^{(d)}}{\sqrt{d}+1}\right]\\
&=\left(1+\frac{1}{(\sqrt{d}+1)^2}+\frac{2}{(\sqrt{d}+1)(d-1)}\right)\mathbb{E}\left[\|v^{(\gamma)}\|^2\right]\\
&=\frac{d}{d-1}\cdot \frac{d-1}{d+2}=\frac{d}{d+2},
\end{align}
where we have used \eqref{norm_vgamma} and \eqref{prod_vgamma}.
Recalling the definition of $\bar{v}^{(\gamma)}$ \eqref{def_vgamma}, we have
\begin{align}
\mathbb{E}[\|\bar{v}^{(\gamma)}\|^2]=\frac{d+2}{d}
\mathbb{E}\left[\left\|v^{(\gamma)}-\frac{1}{\sqrt{d}+1}v^{(d)}\right\|^2\right]
=1.
\end{align}

\item We compute the expectation of the inner product of $\bar{v}^{(\gamma)}$ and $\bar{v}^{(\gamma')}$  for $1\leq \gamma < \gamma'\leq d-1$. We have
\begin{align}
\mathbb{E}\left[\left(v^{(\gamma)}-\frac{1}{\sqrt{d}+1}v^{(d)}\right)\cdot\left(v^{(\gamma')}-\frac{1}{\sqrt{d}+1}v^{(d)}\right)\right]
&=\mathbb{E}\left[v^{(\gamma)}\cdot v^{(\gamma')} +\frac{1}{(\sqrt{d}+1)^2}\|v^{(d)}\|^2-\frac{2v^{(\gamma)}\cdot v^{(d)}}{\sqrt{d}+1}\right]\\
&=\left(\frac{-1}{d-1}+\frac{1}{(\sqrt{d}+1)^2}+\frac{2}{(\sqrt{d}+1)(d-1)}\right)\mathbb{E}\left[\|v^{(\gamma)}\|^2\right]\\
&=0,
\end{align}
which leads $\mathbb{E}[\bar{v}^{(\gamma)}\cdot \bar{v}^{(\gamma')}]=0$.

\item It is sufficient that we consider $Var[v\cdot \bar{v}^{(\gamma)}]$ for $v\in \bar{V}$.
We first evaluate $Var[v\cdot \bar{v}^{(\gamma)}]$ for $v\in V$ ($V$ is defined in Lemma \ref{lem_exp_var}).
Letting $a=1/(\sqrt{d}+1)$, we have 
\begin{align}
Var[v\cdot (v^{(\gamma)}-av^{(d)})]&=Var[v\cdot v^{(\gamma)}-a v\cdot v^{(d)}]\\
&=Var[v\cdot v^{(\gamma)}]+ Var[av\cdot v^{(d)}]
-2Cov(v\cdot v^{(\gamma)}, av\cdot v^{(d)}).
\end{align}
By Cauchy–Schwarz inequality, we have
\begin{align}
|Cov(v\cdot v^{(\gamma)}, av\cdot v^{(d)})|\leq \sqrt{Var[v\cdot v^{(\gamma)}]Var[av\cdot v^{(d)}]}.
\end{align}
Then, we have
\begin{align}
Var[v\cdot (v^{(\gamma)}-av^{(d)})]&\leq Var[v\cdot v^{(\gamma)}]+ a^2 Var[v\cdot v^{(d)}]
+2a \sqrt{Var[v\cdot v^{(\gamma)}]Var[v\cdot v^{(d)}]}\\
&\leq (1+a^2 +2a)\frac{C}{m}.
\end{align}
Recalling $\bar{v}^{(\gamma)}=(1+2/d)(v^{(\gamma)}-av^{(d)})$, there is a certain positive constant $\bar{C}_1$ such that 
\begin{align}
\forall v\in V,\ Var[v\cdot \bar{v}^{(\gamma)}]\leq 
\frac{\bar{C}_1}{m}.
\end{align}

Next, we evaluate $Var[\bar{v}^{(\gamma')}\cdot \bar{v}^{(\gamma)}]$ for $\gamma' \leq \gamma$. Note that
\begin{align}
Var[\bar{v}^{(\gamma')}\cdot (v^{(\gamma)}-av^{(d)})]&=Var[\bar{v}^{(\gamma')}\cdot v^{(\gamma)}-a \bar{v}^{(\gamma')}\cdot v^{(d)}]\\
&=Var[\bar{v}^{(\gamma')}\cdot v^{(\gamma)}]+ Var[a\bar{v}^{(\gamma')}\cdot v^{(d)}]
-2Cov(\bar{v}^{(\gamma')}\cdot v^{(\gamma)}, a\bar{v}^{(\gamma')}\cdot v^{(d)})\\
&\leq (1+a^2 +2a)\frac{\bar{C}_1}{m}.
\end{align}
Thus, there is a certain positive constant $\bar{C}_2$ such that 
\begin{align}
Var[\bar{v}^{(\gamma')}\cdot \bar{v}^{(\gamma)}]\leq 
\frac{\bar{C}_2}{m}.
\end{align}
\end{enumerate}

\end{document}